# Getting Ready for the EU AI Act in Healthcare. A call for Sustainable AI Development and Deployment.

John Brandt Brodersen, University of Copenhagen, Denmark & UiT The Arctic University of Norway.
Ilaria Amelia Caggiano, Research Center in European Private Law (ReCEPL),Università degli studi Suor Orsola, Naples, Italy.
Pedro Kringen, Arcada University of Applied Science, Helsinki, Finland.
Vince Istvan Madai, QUEST Centre for Responsible Research, Berlin Institute of Health, Charité - Universitätsmedizin Berlin, Germany.
Walter Osika, Karolinska Institutet, Stockholm, Sweden & Stockholm Health Care Services, Region Stockholm, Stockholm, Sweden
Giovanni Sartor, CIRSFID-Alma AI, University of Bologna, Italy & EUI, Florence, Italy.
Ellen Svensson, Karolinska Institutet, Stockholm, Sweden & Stockholm University, Stockholm, Sweden.
Magnus Westerlund, Arcada University of Applied Science, Helsinki, Finland.
Roberto V. Zicari, Graduate School of Data Science, Seoul National University, South Korea.

On behalf of the Z-Inspection® Initiative.

## Abstract

Assessments of trustworthiness have become a cornerstone of responsible AI development. Especially in high-stakes fields like healthcare, aligning technical, evidence-based, and ethical practices with forthcoming legal requirements is increasingly urgent. We argue that developers and deployers of AI systems for the medical domain should be proactive and take steps to progressively ensure that such systems — both those currently in use and those being developed or planned — respect the requirements of the AI Act, which has come into force in August 2024. This is necessary if full and effective compliance is to be ensured when the most relevant provisions of the Act become effective (August 2026). The engagement with the AI Act cannot be viewed as a formalistic exercise. Compliance with the AI Act needs to be carried out through the proactive commitment to the ethical principles of trustworthy AI. These principles provide the background for the Act, which mentions them several times and connects them to the protection of public interest (in Recital 7). They can be used to interpret and apply the Act's provisions and to identify good practices, increasing the validity and sustainability of AI systems over time.

## The AI Act and the healthcare domain

As assessments of trustworthiness increasingly anchor the discourse on responsible AI, especially in safety-critical sectors like healthcare, there is a growing need for developers, deployers, and users to move beyond abstract principles and begin operationalizing trustworthiness in concrete, auditable ways (Lekardi et al., 2025; Zicari et al., 2022). This involves not only aligning technical performance and clinical utility with ethical and societal considerations but also preparing for the binding legal obligations introduced by the EU AI Act (EU 2024).

The AI Act includes further provisions, which, while not specifically addressing the healthcare domain, may be relevant for medical applications, such as the transparency rules aimed to enable users to distinguish human and AI-generated outcomes as well as the requirements for general-purpose AI models (such as the Large Language Models LLMS, which can be deployed for analysing or generating medical content).

This very complex piece of legislation, stretching over 150 pages, provides the first comprehensive regulation of AI in the world. It entered into force on 1 August 2024, after the EU Parliament and the Council converged on the final text. Its purpose is to improve the functioning of the EU internal market and promote the uptake of human-centred and trustworthy artificial intelligence. The Act's provision, as explicitly affirmed in Recital 7, takes into account the "Ethics guidelines for trustworthy AI of the High-Level Expert Group on Artificial Intelligence (AI HLEG)", which therefore are highly relevant for the correct interpretation and application of the Act. At the same time, the Act aims to ensure a high level of protection of health, safety, and fundamental rights, including democracy, the rule of law, and environmental protection, against the (unintended) harmful effects of AI systems in the EU and supporting innovation. Within healthcare, the AI Act aims to support beneficial uses of AI, promote a patient-centred approach, and ensure human oversight over AI-supported medical interventions, while preventing harmful accidents, false negatives and false positives, under- and overdiagnosis, and unfair treatment of individual patients and groups of them.

The AI Act prohibits certain AI systems (those which are deemed to pose inevitably unacceptable risks), including systems that exploit vulnerable groups, biometric categorisation, social scoring, individual predictive policing, facial recognition systems in public places, emotion recognition systems in work-places and educational institutions, and 'real-time' remote biometric identification systems in publicly accessible space (which are permitted only under strictly defined and exceptional circumstances, such as for the prevention of serious crime or threats to public security, and subject to judicial or administrative authorization).

It requires certain categories of AI systems (those which are deemed to pose high risks) to be subject to a regime meant to minimise risks, so that only systems whose risks are acceptable are deployed in the EU. The regime for high-risk systems includes a set of requirements pertinent to risk management, data and data governance, technical documentation, record keeping, transparency and information to users, human oversight, accuracy, robustness and cybersecurity, quality management, proportionate evaluation of impacts on people's health using relevant prognostic outcomes like morbidity, mortality and quality of of life, safety, fundamental rights and values. Compliance with such requirements will be assessed through a certification procedure.

In the healthcare domain, high-risk applications include, for example, the development of the following applications: a) AI systems intended to evaluate and classify emergency calls by natural persons or to be used to dispatch, or to establish priority in the dispatching of, emergency first response services, including … emergency healthcare patient triage systems (Annex III, item 5); b) Medical devices (consisting of or including AI systems, required to undergo a third-party conformity assessment, according to Regulation (EU) 2017/745 (medical devices) and Regulation (EU) 2017/746 (n vitro diagnostic medical devices)). These include diagnostic, prediction, or screening tools.

The AI Act is not the only EU regulatory instrument dealing with AI in medicine: relevant regulations include the General Data Protection Regulation - GDPR (EU 2016), the Medical Devices Regulations - MRD (EU 2017) and European Health Data Space Regulation - EHDS (EU 2022) (see Table 1, which illustrates *overlaps relevant for compliance in clinical practice).* As shown in Table 1, this invites an effort of integrated compliance, where the requirements established by these instruments are jointly taken into account.

| Regulatory Area | AI Act (2024) | GDPR | MDR (Medical Device Regulation) | EHDS (European Health Data Space) |
|---|---|---|---|---|
| Scope | AI systems (esp. high-risk) | Personal data | Medical devices | Health data use & exchange |
| Main Purpose | Safety, fundamental rights, trustworthiness | Data protection & privacy | Device safety, performance & clinical effectiveness | Health data sharing & interoperability |
| Risk Management | Extensive mandatory processes | Impact assessments required | Extensive, mandatory evaluation & certification | Data governance & security frameworks |
| Transparency & Explainability | Explicit transparency obligations, right to explanation | Right to information | Documentation and user information requirements | Transparency in data use & access |
| Human Oversight & Accountability | Required explicitly | Indirect (data controller accountability) | Manufacturer accountability & notified bodies oversight | Data holder accountability & governance bodies oversight |
| Compliance & Enforcement | Certification, audits, substantial fines | Audits, substantial fines | Notified bodies audits, certification & penalties | Audits, fines & supervisory authorities |

***Table 1.*** *An overview of the EU AI Act compared to existing healthcare regulatory frameworks (GDPR, MDR, EHDS).*

**The schedule of the AI Act and proactive initiatives to be taken**

As noted above, the AI Act came into force on 1 August 2024, with delayed applicability for certain provisions. However, some important parts have already become effective, such as the ban on prohibited systems, effective since 2 February 2025. On 2 August 2025, certain provisions concerning general-purpose AI systems will also become applicable, followed by specific rules regarding governance obligations and related sanctions. The key provisions for the medical domain—i.e., those on high-risk systems—will take effect only on 1 August 2026, exactly two years after the Act entered into force.

This interval between the entry into force and the coming into effect of these provisions has the purpose of enabling developers and deployers to develop methods, skills, and procedures to be able to comply when the provisions become applicable. This requires a **proactive attitude**, namely, to immediately engage in setting up and testing appropriate procedures, and apply them to progressively ensure that high-risk AI systems match the Act requirement. A proactive approach will enable companies and deployers (such as hospitals in the medical domain) to have in place whatever is needed by August 2026, but also to ensure that existing systems, and those being developed, are compliant. It is true that systems put on the market before August 2026 may continue to be used. However, all high-risk systems need to comply with the AI Act if they undergo substantial modifications after 2026, and those used by public administrations are in any case required to be compliant by 2030. Thus, in any case, high-risk systems should be made compliant as early as possible.

The proactive attitude includes, first of all, examining what AI applications in healthcare are being used, developed or planned for, and considering how they fit into the AI Act, and importantly, whether they fall under the class of high-risk systems. Should that be the case, it is necessary to engage without delay in the critical examination of such applications, to check the extent to which they comply with the general requirements of trustworthy AI and with any more specific requirements established by the AI Act. Specifically, in the medical domain, it is critical to consider whether other relevant legal, evidence-based and ethical requirements are fully respected, such as those pertaining to medical ethics, medical law, and data protection, and how they are likely to be applied when AI systems are used. As shown in Table 1 it might be advisable to consider interrelationships between specific requirements described by the AI Act and the already existing regulatory framework implemented in the health sector, such as the General Data Protection Regulation (GDPR), the Medical Device Regulation (MDR), the European Health Data Space (EHDS). A clarification of how potential overlaps or additions should be properly handled would ease implementation, reduce the ambiguity, and facilitate regulatory alignment.

It is fundamental that such an assessment is not undertaken with a merely defensive mindset, e.g., with the purpose of doing the minimum that is required to avoid possible sanctions. It should, rather, be based on the full and proactive endorsement of the values underlying trustworthy AI, such as respect for autonomy, the engagement for producing sustainable, beneficial and not harmful outcomes for all involved individuals and social groups, concern for equity, fairness and non-discrimination, the pursuit of transparency and explainability. This also raises the question about whose interests and values are addressed in these processes. Without a conscious inclusion of perspectives from patients, healthcare professionals from all areas (nurses, physical therapists, etc.), and marginalised groups, the ethical assessment may only reflect the viewpoints of institutional experts. Tensions between interests, rights and values at stake may emerge, for instance, between the use of patients' data for medical research and data protection rights of patients, or between the accuracy of diagnostic systems and their explicability. Such tensions must be recognised and addressed in the attempt to provide a proportionate outcome, which implements the ethical values of trustworthy AI, consistently with the fundamental principles of EU law.

Further, it is worth investigating who ultimately has the final responsibility in clinical decision-making when AI is used. Especially in cases where human control is formally present but, in reality, is limited or superficial. Without a clear line of responsibility, there is a risk of responsibility diffusion, which in turn may erode patients' safety and professional integrity. Reflections on the practical challenges posed by translating high-level abstract ethical values are also warranted. These values must be articulated into concrete procedures, documentation requirements and quality metrics within the specific domain. Without robust methods for translation into practice, trustworthiness might remain a rhetorical ambition rather than a measurable and testable property of AI systems.

## Why perform a Trustworthy AI Assessment before the day the EU AI Act becomes binding?

An approach grounded in the proactive and genuine endorsement of the values underlying trustworthy AI—such as respect for autonomy, equity, transparency, and the commitment to sustainable and beneficial outcomes—could also significantly enhance the overall quality and long-term sustainability of AI systems. Recent research shows that integrating ethics and trustworthiness assessments into AI system development not only addresses ethical concerns but also contributes to the improvement of the systems themselves by enhancing their reliability, validity, transparency, and alignment with societal values. In addition, recent studies assessing real-world AI systems have found that incorporating ethical and trustworthiness evaluations leads to more robust and valid AI applications, although there is still limited empirical evidence available on how these assessments impact long-term system performance or public trust in real-world clinical use.

These assessments help identify potential issues early in the development process, allowing for timely interventions that enhance system performance and trustworthiness, and have been shown to enhance public trust and ensure that AI systems adhere to ethical standards, thereby improving their overall reliability and acceptance (Wirth et al., 2025). One challenge is that a legalistic-formalistic adoption of the AI Act, only focusing on reducing the risk of incurring legal problems, may have two serious drawbacks.

On the one hand, the attempt to exhibit complaint procedures may lead to unnecessary, excessive bureaucratic processes, which may reduce flexibility and proper adaptation to clinical practice. This can also hamper innovation and user agility in acute or complex situations, under institutional pressure. On the other hand, the assessments of all potential unintended harms may be reduced to producing an appearance of compliance, while failing to properly detect, assess, and mitigate those potential harms, minimise the ethical risks, via a superficial "box ticking" exercise. This might in turn lead to superficial assessments and, in worst cases, might reduce rather than increase trustworthiness, sometimes referred to as ethics washing. We should be vigilant to ensure that compliance is effective, rather than an exercise in law whitewashing, as we should be vigilant to ensure that effective ethical trustworthiness is delivered, rather than being substituted by ethics whitewashing.

A passive attitude would give credence to the widespread criticisms regarding the purely limiting effect of EU regulation on technological development. A proactive attitude, and in this sense, an AI Act taken seriously, is suited to give concrete substance to that legislation, developing reliable and time-tested methods to assess the impact of AI systems even in the medium to long term.

Interestingly, an innovative 2024 study introduced a multi-agent system designed to generate ethically sound AI outputs. By structuring AI agents to engage in debates on real-world ethical issues, the system produced more comprehensive and ethically aligned code and documentation compared to baseline models. This demonstrates that embedding ethical assessments into AI development can enhance the ethical quality of AI outputs (de Cerqueira et al., 2024).

**Ethics-based assessments and AI Act compliance: the case of Z-Inspection®**

As an example showcasing the significance of a broad and proactive ethic-based evaluation, in the context of compliance with the AI Act, we can consider the Z-Inspection®-methodology - a framework for evaluating AI trustworthiness- which has already been tested and deployed in several use cases.

The strength in the Z-Inspection®-methodology lies in its resistance to simplifications. It is domain-specific, involves a wide range of different domain specialists and brings to light ethical issues and tensions rather than obscuring these behind a formal tick box compliance.

Research involving the Z-Inspection® process has revealed that systematic assessments help identify ethical risks and improve systemdesign. Applying this process in healthcare AI systems led to better alignment with ethical principles and increased stakeholder trust (Zicari et al., 2022).

We suggest that systematic assessments of trustworthiness, such as the Z-Inspection® process, may set the stage for the more organisational and administrative steps involved in compliance with the Act. Z-Inspection®, by adopting a broad, ethically, socially, evidence-based and technologically oriented perspective, that directly engages with identifying, evaluating and mitigating risks. It includes the critical appraisal of all impacts on the interests, rights and values at stake, and of the resulting tensions and synergies, including continuous post-deployment of ethical performance.

This is especially relevant for high-risk AI systems currently being tested in real life in Europe before the AI Act is legally binding. The early assessment of such systems is in line with the AI Pact, a voluntary initiative started by the Commission to encourage early adoption of the AI Act requirements (European Commission, 2024). The AI Pact is structured around two pillars.

The first pillar deals with "Community Engagement and Knowledge Sharing". It fosters a collaborative community, open to all stakeholders—including companies, non-profits, academia, and public authorities. Participants share experiences, best practices, and internal policies related to AI governance and compliance. The AI Office organises webinars and workshops to provide insights into the AI Act and gather feedback on implementation challenges.

The second Pillar focuses on "Voluntary Pledges for Early Compliance", i.e., on inviting organisations to make voluntary commitments to proactively implement certain AI Act measures ahead of their legal applicability. Participants are required to develop an AI governance strategy to guide responsible AI adoption within the organisation. Identifying and mapping AI systems that may be classified as high-risk under the AI Act. They also commit to promoting AI awareness and literacy among staff to ensure ethical and informed use of AI technologies. Until now, over 200 companies, including major tech firms and SMEs across various sectors, have signed the AI Pact pledges. These organisations commit to reporting on their progress 12 months after signing, enhancing transparency and accountability in AI development and deployment.

**Conclusion**

Here we synthesise the outcome of the above discussion, and provide some direction for future work, to ensure the best early implementation of the AI Act in healthcare, through the synergy of ethical and legal requirements. The key takeaways from the above discussions are the following:

- The EU AI Act is now in force, with provisions on high-risk systems—especially relevant to healthcare—coming into effect on 1 August 2026. These will require compliance with detailed obligations around risk management, transparency, and ethical alignment.
- High-risk AI systems in healthcare include emergency triage tools, AI-driven medical diagnostics, prediction and screening tools, and further AI-enabled medical devices. These must undergo rigorous certification and documentation procedures to ensure that they improve people's and patients' prognosis, safety, and respect for fundamental rights.
- Proactive, value-driven compliance is encouraged, not just formalistic "box-ticking." Trustworthiness assessments should be grounded in ethical principles like autonomy, fairness, transparency, and sustainability.
- The AI Pact, launched by the European Commission, encourages early compliance through voluntary pledges and knowledge sharing. It helps organisations prepare for full enforcement through a combination of governance strategy, system mapping, and staff education.
- Ethical and trustworthiness assessments, such as the Z-Inspection® process, can improve system design and public trust while preventing shallow or ineffective compliance practices that stifle innovation or overlook real ethical risks.

On this basis, we suggest the following future directions. Before August 2026, it is necessary to get ready for the full effect of the AI Act and to ensure that AI applications are legally compliant and trustworthy:

- Map existing and planned AI systems to determine which qualify as "high-risk" under the EU AI Act.
- Initiate internal trustworthy AI assessments for each high-risk system, using established methodologies (e.g., Z-Inspection®).
- Develop an AI governance strategy tailored to your organisation's healthcare context, aligned with both ethical standards and legal requirements.
- Join the AI Pact or similar initiatives to benefit from peer learning, support, and visibility in compliance efforts.
- Engage cross-disciplinary expertise (technical, legal, clinical, ethical) to evaluate systems from multiple angles and prepare documentation for conformity assessment.
- Start staff training to build AI literacy and understanding of ethical implications in practice.

Beyond August 2026, it is necessary to ensure that compliance is ensured over time, while working to improve trustworthiness:

- Maintain ongoing monitoring and reassessment of AI systems, especially after modifications, to ensure continued compliance and responsiveness to new risks.
- Integrate feedback mechanisms from users, clinicians, and patients into the evaluation of AI systems.
- Contribute to evidence generation around ethical assessments' impact on clinical outcomes, public trust, and system performance.
- Collaborate in ecosystem-wide efforts to standardise trustworthy AI practices across institutions and vendors.
- Work towards alignment with 2030 requirements, especially for AI systems used by public administrations, which must be fully compliant by then, regardless of deployment date.

A meaningful implementation of the AI Act in healthcare calls for more than legalistic or technical compliance - it also requires ethically grounded stakeholder approaches that set professional responsibility and human dignity at the center. By addressing this proactively, the healthcare sector has a unique opportunity to take a lead in the development of AI that is compliant with the law, trustworthy, and beneficial to all.

## Legislation

EU (2016) GDPR (General Data Protection Regulation). European Union. (2016). Regulation (EU) 2016/679 of the European Parliament and of the Council of 27 April 2016 on the protection of natural persons with regard to the processing of personal data and on the free movement of such data (General Data Protection Regulation). Official Journal of the European Union. https://eur-lex.europa.eu/eli/reg/2016/679/oj

EU (2017). MDR (Medical Device Regulation)European Union. (2017). Regulation (EU) 2017/745 of the European Parliament and of the Council of 5 April 2017 on medical devices, amending Directive 2001/83/EC, Regulation (EC) No 178/2002 and Regulation (EC) No 1223/2009 and repealing Council Directives 90/385/EEC and 93/42/EEC (Medical Device Regulation). Official Journal of the European Union. https://eur-lex.europa.eu/eli/reg/2017/745/oj

**Grants**
Roberto V. Zicari is supported by the BrainPool Program through the National Research Foundation of Korea (NRF) funded by the Ministry of Science and ICT (grant number: 2022H1D3A2A01082266, Research Title: Assessing Trustworthy AI).
Giovanni Sartor is supported by the ERC-Advanced CompuLaw project, under Horizon 2020 (Grant Agreement N. 833647).
Walter Osika is supported by grants from the Swedish Research Council Formas: Mind4Change (nr 2019-00390),  TransVision (nr 2019-01969), and Playbook for New Health and New Urban Patterns (nr 2023-02391).